# Violation-Informed Spatio-Temporal Adaptive Targeting Framework for EV-Driven Distribution System Expansion Planning

Linhan Fang, *Student Member, IEEE* and Xingpeng Li, *Senior Member, IEEE*

***Abstract*—The rapid adoption of electric vehicles (EVs) can cause severe voltage drops and line current overloads in distribution networks, creating an urgent need for scalable expansion planning methods. This paper proposes a computationally efficient violation-informed spatio-temporal adaptive targeting (STAT) framework for EV-driven distribution system expansion planning. The framework first identifies potential voltage and current violations through a violation analysis model, and then mitigates them through a joint optimal expansion planning model that co-optimizes investment decisions for line reconductoring, shunt capacitors, and battery energy storage systems. To reduce computational burden, the proposed STAT-temporal criticality assessment (STAT-TCA) method extracts primitive stress events from annual operating data, derives an initial set of candidate planning horizons from signature-consistent segments, and selects a final transferable critical horizon set through cross-horizon validation based on optimization feasibility and cost. Meanwhile, the proposed STAT-adaptive spatial targeting (STAT-AST) method constructs device-specific spatial features for BESS and SC siting to retain compact yet high-impact candidate bus sets. Case studies on 33-bus and 240-bus distribution systems demonstrate that the proposed STAT framework can substantially reduce the temporal and spatial planning dimensions while preserving planning fidelity. Full-year validation further confirms that the resulting investment plans can eliminate EV-induced voltage and thermal violations while maintaining feasible BESS operations.**

## NOMENCLATURE

*A. Indices and Sets*

| | |
|---|---|
| $g \in \mathcal{G}$ | Index and set of substation buses |
| $i \in \mathcal{N}$ | Index and set of non-substation buses |
| $(i,j) \in \mathcal{L}$ | Index and set of branch information |
| $\mathcal{L}_{cand}$ | Set of upgrade candidate lines branch information |
| $t \in \mathcal{T}$ | Index and set of time intervals |
| $j \in \mathcal{D}(i)$ | Index and set of downstream nodes of node $i$ |
| $k \in \mathcal{U}(i)$ | Index and set of upstream nodes of node $i$ |
| $b \in \mathcal{B}$ | Index and set of candidate BESS installation buses |
| $s \in \mathcal{S}$ | Index and set of candidate SC installation buses |
| $o \in \mathcal{O}_{ij}$ | Index and set of upgraded cable candidate types |

*B. Parameters*

| | |
|---|---|
| $r_{ki}$ | Resistance of the line connecting node $k$ and node $i$ |
| $x_{ki}$ | Reactance of the line connecting node $k$ and node $i$ |
| $P_{\text{load},i}^{t}$ | Active power demand of node $i$ at time $t$ |
| $Q_{\text{load},i}^{t}$ | Reactive power demand of node $i$ at time $t$ |
| $I_{ij}^{\text{lim}}$ | Maximum current ampacity between node $i$ and $j$ |
| $C_{ij}^{o}$ | The cost of upgrading line between node $i$ and $j$ with the $o$-th cable type |
| $N_{\text{SC}}^{\max}$ | Maximum number of the installed SC |
| $C_{\text{SC}}^{\text{fix}}, C_{\text{SC}}^{\text{step}}$ | The fixed installation cost / capacity cost of SC |
| $E_{\min}^{\text{cap}}, E_{\max}^{\text{cap}}$ | Minimum / maximum capacity of installed BESS |
| $SoC^{\min}$ | Minimum limit of state of charge of BESS |
| $SoC^{\max}$ | Maximum limit of state of charge of BESS |
| $\eta_{\text{ch}}, \eta_{\text{dis}}$ | Charging / Discharging power efficiency |
| $v^{\min}, v^{\max}$ | Voltage lower limit / upper limit |
| $C_{\text{rate}}^{\text{ch}}, C_{\text{rate}}^{\text{dis}}$ | Charge / discharge C-rate of BESS |
| $K_{\text{inv}}$ | Inverter capacity scaling factor |
| $C_{\text{BESS}}^{\text{eng}}, C_{\text{BESS}}^{\text{pow}}$ | The energy-related capacity cost for battery packs / power-related rating cost for inverters |

*C. Variables*

| | |
|---|---|
| $P_{ij}^{t}, Q_{ij}^{t}$ | Power flow from node $i$ to node $j$ at time $t$ |
| $P_{ij}^{\text{loss},t}$ | Branch power loss between node $i$ and $j$ |
| $Q_{ij}^{\text{loss},t}$ | Branch reactive power loss between node $i$ and $j$ |
| $v_{i}^{t}$ | The square of voltage magnitude of node $i$ at time $t$ |
| $l_{ij}^{t}$ | The square of the current in the line at time $t$ |
| $P_{b,t}^{\text{BESS_ch}}$ | Active charging power of BESS at bus $b$ at time $t$ |
| $P_{b,t}^{\text{BESS_dis}}$ | Active discharging power of BESS at bus $b$ at time $t$ |
| $Q_{b,t}^{\text{BESS_inj}}$ | Reactive power flowing from BESS into grid at bus $b$ at time $t$ |
| $Q_{b,t}^{\text{BESS_abs}}$ | Reactive power flowing from grid into BESS at bus $b$ at time $t$ |
| $SoC_{b,t}$ | The state of charge of BESS at bus $b$ at time $t$ |
| $E_{b}^{\text{cap}}$ | The capacity of BESS at bus $b$ |
| $E_{b,t}^{\text{ess}}$ | The amount of the stored energy of BESS of the bus $b$ at time $t$ |
| $S_{\text{inv},b}$ | The apparent power of the inverter of the bus $b$ |
| $\mathfrak{s}_{i,t}^{V}, \mathfrak{s}_{ij,t}^{I}$ | The severity of voltage / current violations at time $t$ |
| $S_{n}^{VP}, S_{n}^{VQ}$ | The voltage sensitivity indices associated with active / reactive power injection at bus $n$ |
| $S_{n}^{IP}$ | The branch current sensitivity index associated with active power injections at bus |

Linhan Fang and Xingpeng Li are with the Department of Electrical and Computer Engineering, University of Houston, Houston, TX, 77204, USA. (e-mail: lfang7@uh.edu; xli83@central.uh.edu).

Acknowledgment: We thank Center Point Energy for providing real smart meter data and for their feedback on method for extracting EV charging loads.

## I. INTRODUCTION

As a critical response to mounting environmental concerns and carbon emissions, electric vehicles (EVs) are emerging as an important cornerstone of modern energy and transportation systems [1]-[3]. Their dual role in reducing vehicular emissions and providing grid-balancing services is crucial for fostering a sustainable energy ecosystem, particularly in synergy with renewable power generation [4]-[6]. In the residential EV charging domain, Level-2 chargers represent a practical common tradeoff between the high upfront investment for DC fast chargers and the limited throughput of Level-1 systems [7]-[8]. Meanwhile, this surge in EV charging loads can greatly stress distribution networks, causing voltage violations and component overloading [9]-[10]. Therefore, reliable distribution network expansion planning is of paramount importance for addressing these challenges and supporting the widespread adoption of EVs.

To address voltage and current violations in EV-rich distribution networks, prior studies have considered both network reinforcement and flexible resource planning. Wang et al. formulated active distribution network expansion planning with multiple distributed energy resources and an EV sharing system, showing that distribution planning can coordinate network investment with distributed resources and charging demand [11]. Other studies have focused on battery energy storage system (BESS) siting and sizing, where storage placement is optimized to improve distribution network operation and reliability under unbalanced operating conditions [12]. Reactive power and voltage regulating devices have also been widely studied for voltage support, especially in distribution systems with high renewable penetration [13]. In addition, data-driven voltage regulation methods use measurement data and distributed optimization to respond to changing operating conditions without relying entirely on detailed network models [14]. These studies show that voltage and current violations can be mitigated through different resources and solution methods, but the coordinated panning of feeder reinforcement, BESS, and reactive power compensation remains essential for capturing their cost and operational trade-offs.

Recent planning studies have shown a clear trend toward integrated resource planning in active distribution networks. Shen et al. incorporated both centralized and distributed energy storage into active distribution network expansion planning, showing that storage can defer conventional network investments when its siting and sizing are optimized together with grid expansion decisions [15]. Yi et al. further co-optimized energy storage allocation and line reinforcement to improve the dispatchability of active distribution networks, and used decomposition to improve convergence of the resulting planning problem [16]. Li et al. considered dynamic thermal ratings of underground cables and transformers in active distribution network expansion planning, highlighting the value of representing infrastructure thermal flexibility in planning models [17]. Yao et al. recently developed a joint planning framework for distribution systems and EV charging stations with PV-grid-EV transactions, reflecting the growing need to couple network investment with EV related flexibility [18]. These studies indicate that distribution planning is moving toward more integrated formulations. However, when feeder reinforcement, reactive compensation, and BESS are planned simultaneously, the model becomes computationally challenging due to mixed investment decisions, time-coupled storage operation, and network security constraints [19]-[20].

Temporal representation is another critical issue in planning models with storage and network constraints. A full chronological horizon can preserve the operating sequence, but it often becomes computationally expensive when investment variables are coupled with hourly dispatch and state of charge (SoC) constraints. To reduce this burden, representative days or weeks have been widely used in power system planning, where a limited number of periods are selected to approximate the annual load and renewable profiles [21]. For storage-related planning, enhanced representative days and system states have been proposed to better preserve chronological effects that are important for energy shifting [22]. Other studies further incorporate extreme net-load days into representative period selection, since typical periods alone may underestimate capacity requirements during highly stressful conditions [23]. More recently, problem-driven scenario reduction has shown that scenario selection should not only approximate the statistical distribution of data, but also reflect the structure and optimality of the target optimization problem [24]. Inspired by this idea, this paper selects planning horizons according to violation-driving stress patterns that influence the final planning decision rather than relying only on peak load or purely statistical representative periods.

Spatial reduction is commonly used to limit the number of candidate locations in distribution planning. Existing studies have used voltage sensitivity to identify effective BESS locations, since storage installed at highly sensitive buses can provide stronger voltage support [25]. Other approaches rely on violation information, such as the duration and intensity of thermal and voltage violation events, to form candidate pools for line upgrades, voltage regulation devices, and storage installations [26]. These methods suggest that candidate selection should reflect both network sensitivity and violation severity. Following this idea, this paper screens candidate nodes using stress-weighted voltage and current sensitivities, electrical distance to stressed regions, local load conditions, and the downstream reactive power support range.

This paper proposes a violation-informed spatio-temporal adaptive targeting (STAT) framework built on a joint optimal expansion planning (OEP) model for mitigating EV-induced voltage and current violations in distribution networks. The proposed OEP model jointly optimizes line reconductoring (LR), BESS siting and sizing, and shunt capacitor (SC) installation, thereby coordinating feeder capacity enhancement, active power support, and reactive power compensation. To address the computational burden of this joint formulation, the proposed STAT framework reduces both the temporal and spatial planning dimensions. The main contributions are summarized as follows:

- A joint OEP model is developed to mitigate EV-induced voltage and current violations in distribution networks by co-optimizing investments in LR, BESS, and SC.
- A STAT-temporal criticality assessment (STAT-TCA) method is developed to extract primitive stress events, derive an initial set of candidate critical horizons from signature-consistent segments, and select a final transferable planning horizon set through cross-horizon feasibility validation rather than conventional statistical representative-period selection.
- A STAT-adaptive spatial targeting (STAT-AST) method is proposed to reduce the candidate node space using device-specific features based on stress-weighted sensitivities, local load characteristics, electrical distance, and downstream reactive demand before solving the planning model.
- The dimension-reduced planning framework is assessed through temporal and spatial benchmarks, a

comprehensive Full-node comparison in the 33-bus system, and full-year fixed-investment validation in both 33-bus and 240-bus systems, demonstrating planning fidelity and larger-system applicability under the tested EV scenarios.

## II. EV-Driven Distribution Expansion Planning Model

### A. Violation Analysis (VA) Model

A VA model is developed in this section to determine whether investment is required for grid reliability purpose by identifying and analyzing the potential voltage and current violation issues under future operating scenarios with load growth. The distribution-flow model is based on the assumption of a radial distribution network and establishes a nonlinear but accurate relationship among nodal voltages, branch power flows, and line currents [27]. The VA model implemented in this paper has an hourly temporal resolution.

The objective of the proposed VA model is to minimize the total active power losses which can be defined as:

$$\text{objective} = \min\left(\sum_{(i,j)\in\mathcal{L}}\sum_{t\in\mathcal{T}} r_{ij}\cdot l_{ij}^{t}\right) \tag{1}$$

where $\mathcal{L}$ is the set of branches and $r_{ij}$ represents the resistance of the line connecting node $i$ and node $j$.

Nodal active and reactive power balance equations at non-substation and substation buses are expressed as (2)-(3) respectively.

$$\begin{cases}\sum_{k\in\mathcal{U}(i)}(P_{ki}^{t}-r_{ki}\cdot l_{ki}^{t})=\sum_{j\in\mathcal{D}(i)}P_{ij}^{t}+P_{\text{load},i}^{t}\\ \sum_{k\in\mathcal{U}(i)}(Q_{ki}^{t}-x_{ki}\cdot l_{ki}^{t})=\sum_{j\in\mathcal{D}(i)}Q_{ij}^{t}+Q_{\text{load},i}^{t}\\ ,\forall i\in\mathcal{N},\forall t\in\mathcal{T}\end{cases} \tag{2}$$

$$\begin{cases}\sum_{k\in\mathcal{U}(g)}(P_{kg}^{t}-r_{kg}\cdot l_{kg}^{t})=-P_{g}^{t}+\sum_{j\in\mathcal{D}(g)}P_{gj}^{t}+P_{\text{load},g}^{t}\\ \sum_{k\in\mathcal{U}(g)}(Q_{kg}^{t}-x_{kg}\cdot l_{kg}^{t})=-Q_{g}^{t}+\sum_{j\in\mathcal{D}(g)}Q_{gj}^{t}+Q_{\text{load},g}^{t}\\ ,\forall g\in\mathcal{G},\forall t\in\mathcal{T}\end{cases} \tag{3}$$

where for each node $i\in\mathcal{N}$, $\mathcal{D}(i)$ and $\mathcal{U}(i)$ represent the sets of its downstream and upstream nodes, respectively.

Voltage drop equation can be obtained as (4) and the constraint between current and power (second-order cone constraint [28]) is defined in (5) and the substation bus voltage $v_g^t$ is set to the annual mean voltage magnitude at the feeder head obtained from the base-case power flow results.

$$v_i^t - v_j^t = 2\left(r_{ij}P_{ij}^t + x_{ij}Q_{ij}^t\right) - \left(r_{ij}^2 + x_{ij}^2\right)l_{ij}^t, \forall (i,j)\in\mathcal{L}, \forall t\in\mathcal{T} \tag{4}$$

$$v_i^t\cdot l_{ij}^t \geq (P_{ij}^t)^2 + (Q_{ij}^t)^2, \forall (i,j)\in\mathcal{L}, \forall t\in\mathcal{T} \tag{5}$$

The relaxed constraints of voltage and current are shown in (6)-(7), and $v^{\max}$ is the square of maximum per-unit voltage between node $i$ and $j$.

$$v_i^t \leq v^{\max}, \forall i\in\mathcal{N}, \forall t\in\mathcal{T} \tag{6}$$

$$0 \leq l_{ij}^t, \forall (i,j)\in\mathcal{L}, \forall t\in\mathcal{T} \tag{7}$$

The resulting voltage and current violation data obtained with the VA model are subsequently processed to select the critical period through the proposed STAT-TCA method and the candidate nodes for BESS and SC installation through the proposed STAT-AST method, as detailed in Section III.

### B. Optimal Expansion Planning (OEP) Model

To mitigate the identified violation issues, the OEP model is developed to co-optimize lines reconductoring and BESS and SC installation. The network and configuration constraints for lines reconductoring and BESS and SC are shown below.

#### B.1 Line Reconductoring Constraints

Built upon the VA model, the original system parameters impedance $r_{ij}$, $x_{ij}$ and ampacity $I_{ij}^{\text{lim}}$ of the overloaded lines are converted into decision variables $r_{ij}^o$, $x_{ij}^o$ and $\bar{I}_{ij}^o$. A binary variable $z_{ij}^o$ is introduced to denote the selection of the $o$-th cable type for line between node $i$ and $j$.

$$\sum_{o\in\mathcal{O}_{ij}} z_{ij}^o = 1, \forall (i,j)\in\mathcal{L}_{\text{cand}} \tag{8}$$

where $\mathcal{L}_{\text{cand}}$ is the candidate set of upgraded lines, and $\mathcal{O}_{ij}$ is the set of upgraded cable types including the existing cable type that can be selected between node $i$ and $j$.

The line current ampacity constraints change dynamically according to the selected upgraded cable as shown in (9).

$$0 \leq l_{ij}^t \leq \sum_{o\in\mathcal{O}_{ij}} (\bar{I}_{ij}^o)^2 z_{ij}^o, \forall (i,j)\in\mathcal{L}_{cand}, \forall t\in\mathcal{T} \tag{9}$$

To maintain the convexity of the optimization model and avoid nonlinear bilinear terms, the Big-M method will be used to reconstruct the network loss and voltage drop constraints and the associated Big-M values are derived from the maximum admissible voltage range, the largest candidate line impedance, and the corresponding ampacity limits.

$$\begin{cases}-M\left(1-z_{ij}^o\right) \leq v_i^t - v_j^t - 2\left(r_{ij}^oP_{ij}^t + x_{ij}^oQ_{ij}^t\right) + \left({r_{ij}^o}^2 + {x_{ij}^o}^2\right)l_{ij}^t\\ \qquad \leq M\left(1-z_{ij}^o\right), \forall (i,j)\in\mathcal{L}_{\text{cand}}, \forall o\in\mathcal{O}_{ij}, \forall t\in\mathcal{T}\\ v_i^t - v_j^t = 2\left(r_{ij}P_{ij}^t + x_{ij}Q_{ij}^t\right) - \left(r_{ij}^2 + x_{ij}^2\right)l_{ij}^t, \forall (i,j)\in\mathcal{L}\backslash\mathcal{L}_{\text{cand}}\end{cases} \tag{10}$$

$$\begin{cases}-M\left(1-z_{ij}^o\right) \leq P_{ij}^{\text{loss},t} - r_{ij}^o l_{ij}^t \leq M\left(1-z_{ij}^o\right), \forall (i,j)\in\mathcal{L}_{\text{cand}}\\ P_{ij}^{\text{loss},t} = r_{ij}\cdot l_{ij}^t, \forall (i,j)\in\mathcal{L}\backslash\mathcal{L}_{\text{cand}}, \forall t\in\mathcal{T}\end{cases} \tag{11}$$

$$\begin{cases}-M\left(1-z_{ij}^o\right) \leq Q_{ij}^{\text{loss},t} - x_{ij}^o l_{ij}^t \leq M\left(1-z_{ij}^o\right), \forall (i,j)\in\mathcal{L}_{\text{cand}}\\ Q_{ij}^{\text{loss},t} = x_{ij}\cdot l_{ij}^t, \forall (i,j)\in\mathcal{L}\backslash\mathcal{L}_{\text{cand}}, \forall t\in\mathcal{T}\end{cases} \tag{12}$$

where $M$ is a positive number and to avoid the generation of non-convex bilinear terms resulting from the multiplication of discrete impedance variables with continuous power flow or line capacity variables, here introduces continuous auxiliary variables $P_{ij}^{loss,t}$ and $Q_{ij}^{loss,t}$ to address the branch losses within constraints (10)-(12). By incorporating the Big-M method, the nonlinear multiplications are precisely linearized into a set of mixed-integer linear constraints. And the strict constraints on voltage and current are defined as (13)-(14):

$$v^{\min} \leq v_i^t \leq v^{\max}, \forall i\in\mathcal{N}, \forall t\in\mathcal{T} \tag{13}$$

$$0 \leq l_{ij}^t \leq (I_{ij}^{\text{lim}})^2, \forall (i,j)\in\mathcal{L}\backslash\mathcal{L}_{\text{cand}}, \forall t\in\mathcal{T} \tag{14}$$

The investment cost for line reconductoring is defined as:

$$C_{\text{Line}}^{\text{inv}} = \sum_{(i,j)\in\mathcal{L}_{\text{cand}}}\sum_{o\in\mathcal{O}_{ij}} (C_{ij}^o z_{ij}^o) \tag{15}$$

where $C_{ij}^o$ represents the cost required to upgrade line between node $i$ and $j$ with the $o$-th cable type.

#### B.2 Shunt Capacitor Constraints

The installation quantity and capacity constraints of SC are as defined in (16)-(17):

$$n_s^{\text{SC}} \in \mathbb{Z}_+, 0 \leq n_s^{\text{SC}} \leq N_{\text{SC}}^{\max}\cdot x_s^{\text{SC}}, \forall s\in\mathcal{S} \tag{16}$$

$$0 \leq q_{s,t}^{\text{SC}} \leq n_s^{\text{SC}}\cdot Q_{\text{step}}^{\text{SC}}, \forall s\in\mathcal{S}, \forall t\in\mathcal{T} \tag{17}$$

where the variable $x_s^{\text{SC}}$ determines the installation status and $n_s^{SC}$ represents the integer number of installed SC units at node $s$. $Q_{\text{step}}^{\text{SC}}$ denotes the SC capacity per installation step and $q_{s,t}^{\text{SC}}$ denotes the hourly reactive output, which is continuously relaxed within the installed capacity during planning stage and discrete output during operational stage.

The investment cost for SC installation is defined as:

$$C_{\mathrm{SC}}^{\mathrm{inv}} = \sum\nolimits_{\forall s \in \mathcal{S}} \left( C_{\mathrm{SC}}^{\mathrm{fix}} x_s^{\mathrm{SC}} + C_{\mathrm{SC}}^{\mathrm{step}} n_s^{\mathrm{SC}} \right) \tag{18}$$

where $C_{\mathrm{SC}}^{\mathrm{fix}}$ and $C_{\mathrm{SC}}^{\mathrm{step}}$ represent the fixed installation cost and capacity cost of the SC, respectively.

*B.3 BESS Operation and Sizing Constraints*

As enforced in (19)-(22), $z_b$ is a binary variable defining the BESS installation status at bus $b$. $u_{b,t}^{\mathrm{ch}}$ and $u_{b,t}^{\mathrm{dis}}$ are binary variables indicating charging and discharging status. $u_{b,t}^{\mathrm{inj}}$ and $u_{b,t}^{\mathrm{abs}}$ are binary variables indicating injection and absorption status respectively. It is not allowed to charge and discharge active power or inject and absorb reactive power to the BESS at the same time. Besides, $E_b^{cap}$ has the upper and lower limit.

$$z_b, u_{b,t}^{\mathrm{inj}}, u_{b,t}^{\mathrm{abs}}, u_{b,t}^{\mathrm{ch}}, u_{b,t}^{\mathrm{dis}} \in \{0,1\}, \forall b \in \mathcal{B}, \forall t \in \mathcal{T} \tag{19}$$

$$z_b \cdot E_{\min}^{\mathrm{cap}} \le E_b^{\mathrm{cap}} \le z_b \cdot E_{\max}^{\mathrm{cap}}, \forall b \in \mathcal{B} \tag{20}$$

$$u_{b,t}^{\mathrm{ch}} + u_{b,t}^{\mathrm{dis}} \le 1, \forall b \in \mathcal{B}, \forall t \in \mathcal{T} \tag{21}$$

$$u_{b,t}^{\mathrm{inj}} + u_{b,t}^{\mathrm{abs}} \le 1, \forall b \in \mathcal{B}, \forall t \in \mathcal{T} \tag{22}$$

To avoid the quadratic constraints as much as possible, we use the Big-M method to define linear power constraints (23)-(28). And the Big-M constants are bounded by the maximum installable BESS energy capacity, the C-rate, and the inverter sizing factor.

$$0 \le P_{b,t}^{\mathrm{BESS_ch}} \le C_{\mathrm{rate}}^{\mathrm{ch}} \cdot E_b^{\mathrm{cap}}, \forall b \in \mathcal{B}, \forall t \in \mathcal{T} \tag{23}$$

$$0 \le P_{b,t}^{\mathrm{BESS_ch}} \le M \cdot u_{b,t}^{\mathrm{ch}}, \forall b \in \mathcal{B}, \forall t \in \mathcal{T} \tag{24}$$

$$0 \le P_{b,t}^{\mathrm{BESS_dis}} \le C_{\mathrm{rate}}^{\mathrm{dis}} \cdot E_b^{\mathrm{cap}}, \forall b \in \mathcal{B}, \forall t \in \mathcal{T} \tag{25}$$

$$0 \le P_{b,t}^{\mathrm{BESS_dis}} \le M \cdot u_{b,t}^{\mathrm{dis}}, \forall b \in \mathcal{B}, \forall t \in \mathcal{T} \tag{26}$$

$$0 \le Q_{b,t}^{\mathrm{BESS_inj}} \le M \cdot u_{b,t}^{\mathrm{inj}}, \forall b \in \mathcal{B}, \forall t \in \mathcal{T} \tag{27}$$

$$0 \le Q_{b,t}^{\mathrm{BESS_abs}} \le M \cdot u_{b,t}^{\mathrm{abs}}, \forall b \in \mathcal{B}, \forall t \in \mathcal{T} \tag{28}$$

The inverter capacity of BESS is defined in (29). The relationship between active power, reactive power, and the apparent power of the inverter is given by (30).

$$S_{\mathrm{inv},b} = K_{\mathrm{inv}} \cdot C_{\mathrm{rate}}^{\mathrm{dis}} \cdot E_b^{\mathrm{cap}}, \forall b \in \mathcal{B} \tag{29}$$

$$\left(P_{b,t}^{\mathrm{BESS_ch}} + P_{b,t}^{\mathrm{BESS_dis}}\right)^2 + \left(Q_{b,t}^{\mathrm{BESS_abs}} + Q_{b,t}^{\mathrm{BESS_inj}}\right)^2 \le S_{\mathrm{inv},b}^2 \tag{30}$$

The constraint on the BESS capacity is shown in (31). And the SoC of the BESS is temporally coupled, where the SoC at any given time depends on its previous state and the charging or discharging actions taken in (32). To ensure cyclical operation, the model constrains the SoC at the beginning time $t_0$ and end time $t_{\mathrm{end}}$ of the optimization period to be equal.

$$SoC^{\min} \cdot E_b^{\mathrm{cap}} \le E_{b,t}^{\mathrm{ess}} \le SoC^{\max} \cdot E_b^{\mathrm{cap}}, \forall b \in \mathcal{B}, \forall t \in \mathcal{T} \tag{31}$$

$$E_{b,t}^{\mathrm{ess}} = E_{b,t-1}^{\mathrm{ess}} + \left(P_{b,t}^{\mathrm{BESS_ch}} \eta_{\mathrm{ch}} - P_{b,t}^{\mathrm{BESS_dis}} / \eta_{\mathrm{dis}}\right) \cdot \Delta t, \forall b \in \mathcal{B} \tag{32}$$

$$E_{b,t_0}^{\mathrm{ess}} = E_{b,t_{\mathrm{end}}}^{\mathrm{ess}}, \forall b \in \mathcal{B} \tag{33}$$

The BESS investment cost is decoupled into energy-related capacity cost $C_{\mathrm{BESS}}^{\mathrm{eng}}$ for battery packs and power-related rating cost $C_{\mathrm{BESS}}^{\mathrm{pow}}$ for inverters. Specifically, the required power capacity is mathematically linked to the installed energy capacity $E_b^{\mathrm{cap}}$ via the specified discharge C-rate. The investment cost for BESS installation is defined as:

$$C_{\mathrm{BESS}}^{\mathrm{inv}} = \sum\nolimits_{b \in \mathcal{B}} \left( C_{\mathrm{BESS}}^{\mathrm{eng}} E_b^{\mathrm{cap}} + C_{\mathrm{BESS}}^{\mathrm{pow}} K_{\mathrm{inv}} C_{\mathrm{rate}}^{\mathrm{dis}} E_b^{\mathrm{cap}} \right) \tag{34}$$

*B.4 Network Constraints and Objective*

The nodal active and reactive power balance equations for non-substation nodes are defined in (35)-(36), while the balance equations for substation node are defined in (37).

$$\sum\nolimits_{k \in \mathcal{U}(i)} \left(P_{ki}^t - P_{ki}^{\mathrm{loss},t}\right) = \sum\nolimits_{j \in \mathcal{D}(i)} P_{ij}^t + P_{\mathrm{load},i}^t + \sum\nolimits_{b \in \mathcal{B}_i} \left(P_{b,t}^{\mathrm{BESS_ch}} - P_{b,t}^{\mathrm{BESS_dis}}\right), \forall i \in \mathcal{N}, \forall t \in \mathcal{T} \tag{35}$$

$$\sum\nolimits_{k \in \mathcal{U}(i)} \left(Q_{ki}^t - Q_{ki}^{\mathrm{loss},t}\right) = \sum\nolimits_{j \in \mathcal{D}(i)} Q_{ij}^t + Q_{\mathrm{load},i}^t + \sum\nolimits_{b \in \mathcal{B}_i} \left(Q_{b,t}^{\mathrm{BESS_abs}} - Q_{b,t}^{\mathrm{BESS_inj}}\right) - \sum\nolimits_{s \in \mathcal{S}_i} q_{s,t}^{\mathrm{SC}}, \forall i \in \mathcal{N}, \forall t \in \mathcal{T} \tag{36}$$

$$\begin{cases} \sum\nolimits_{k \in \mathcal{U}(g)} \left(P_{kg}^t - P_{kg}^{\mathrm{loss},t}\right) = -P_g^t + \sum\nolimits_{j \in \mathcal{D}(g)} P_{gj}^t + P_{\mathrm{load},g}^t \\ \sum\nolimits_{k \in \mathcal{U}(g)} \left(Q_{kg}^t - Q_{kg}^{\mathrm{loss},t}\right) = -Q_g^t + \sum\nolimits_{j \in \mathcal{D}(g)} Q_{gj}^t + Q_{\mathrm{load},g}^t \\ \forall g \in \mathcal{G}, \forall t \in \mathcal{T} \end{cases} \tag{37}$$

where the $\mathcal{B}_i$ and $\mathcal{S}_i$ denote the sets of BESS and SC integrated at node $i$, respectively. The objective function of the OEP model is to minimize the total annualized investment cost of line reconductoring and BESS and SC installation in (39). To ensure a fair economic evaluation across assets with varying operational lifespans, the initial investments are annualized using the capital recovery factor $\gamma_{CRF}$ in (38), where $\tau$ and $Y$ are the discount rate and specific asset lifetime, respectively.

$$\gamma_{\mathrm{CRF}} = \frac{\tau(1+\tau)^Y}{(1+\tau)^Y - 1} \tag{38}$$

$$\text{Objective:} \min\left(\gamma_{\mathrm{CRF}}^{\mathrm{Line}} C_{\mathrm{Line}}^{\mathrm{inv}} + \gamma_{\mathrm{CRF}}^{\mathrm{SC}} C_{\mathrm{SC}}^{\mathrm{inv}} + \gamma_{\mathrm{CRF}}^{\mathrm{BESS}} C_{\mathrm{BESS}}^{\mathrm{inv}}\right) \tag{39}$$

When the OEP model is executed for a full year, many hourly intervals merely inflate the computational burden without impacting the optimal decisions. By applying the proposed STAT framework to reduce the temporal dimension, the derived T-OEP model can achieve results with a significantly lighter computational footprint. And the overall planning formulation still remains a mixed-integer second-order cone programming (MISOCP) model. A comparative summary of the constraints and solution time dimensions for the VA, OEP, and T-OEP models is detailed in Table 1.

Table 1. Comparison of Key Settings between VA, OEP and T-OEP models.

| Model | Objective | Constraints | Timeframe $\mathcal{T}$ |
|---|---|---|---|
| VA model | (1) | (2)-(6) | Full-year planning horizon |
| OEP model | (39) | (5), (8)-(38) | Full-year planning horizon |
| T-OEP model | (39) | (5), (8)-(38) | Critical planning horizon |

Note: For T-OEP model, the cyclic terminal SoC constraint in (33) is relaxed.

## III. The Proposed STAT Framework

The proposed violation-informed spatio-temporal adaptive targeting framework consists of two preprocessing stages that coordinate together to decrease the complexity of the T-OEP model: (i) the STAT-TCA method identifies critical horizons to reduce planning runtime; and (ii) the STAT-AST method pinpoints reduced sets of candidate buses for BESS and SC installation, mitigating the computing risk from an overly large candidate buses pool. As shown in Fig. 1, annual VA model outputs are processed by the proposed STAT-TCA and STAT-AST methods to obtain critical horizons and reduced BESS and SC candidate sets. These reduced temporal and spatial inputs are then used in the T-OEP model to determine LR, BESS, and SC planning decisions, followed by full-year operational validation.

*A. Temporal Criticality Assessment*

The first stage of the STAT framework evaluates temporal criticality by extracting critical high-stress planning horizons from the annual operating trajectory. The annual data are first organized into current-violation, voltage-violation, and load event categories, and each event is characterized by a multidimensional feature vector that captures the magnitude, duration, and stress extent. A Pareto screening procedure is then applied to retain the most critical events, which are subsequently remapped onto the annual timeline to form signature-consistent segments. The final planning horizons are recovered from the retained segments and later validated through cross-horizon feasibility testing.

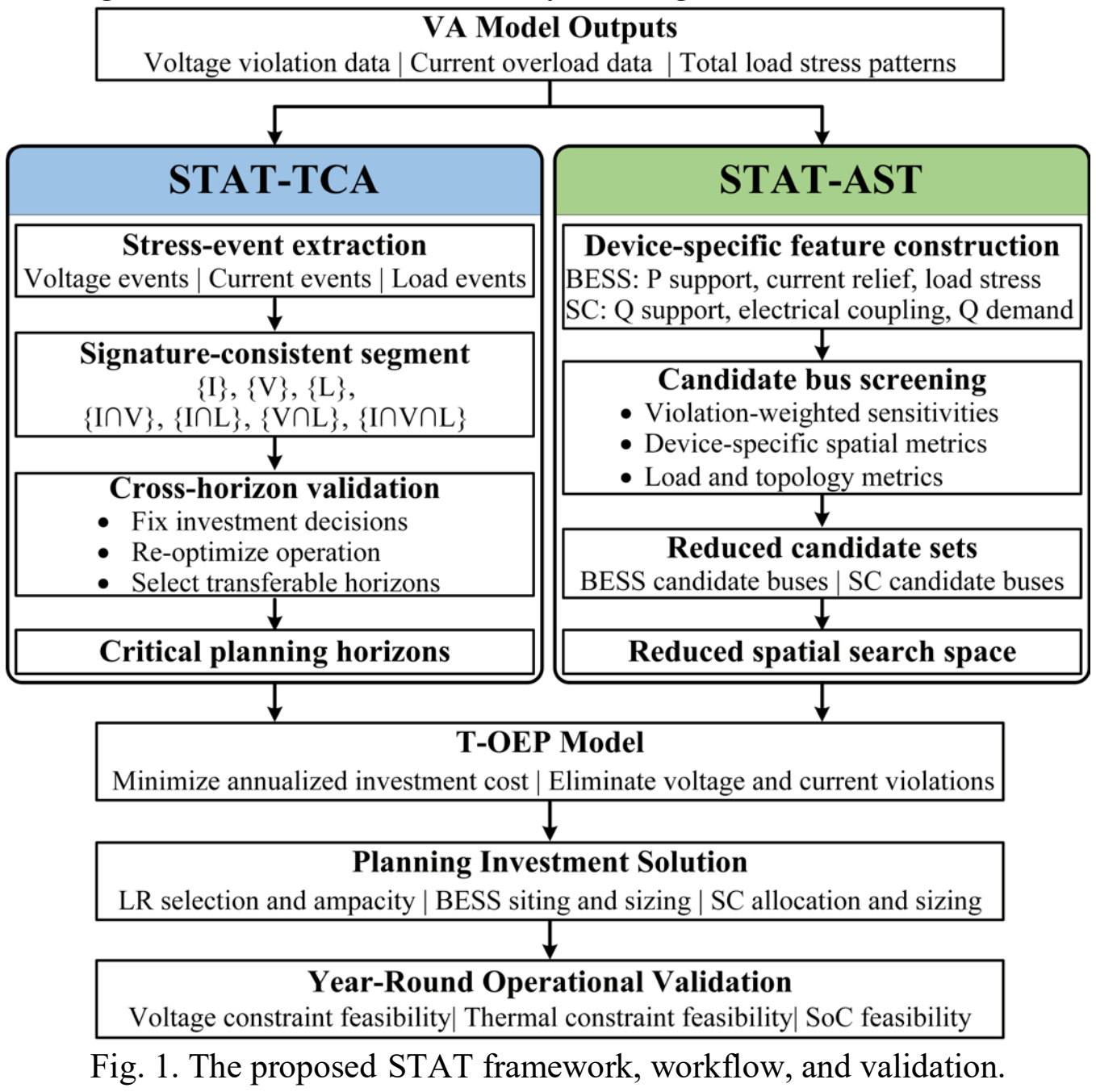

Fig. 1. The proposed STAT framework, workflow, and validation.

### *A.1 Stress-Event Extraction*

The severity of hourly voltage and current violations at node $i$ are defined in (40)-(41):

$$\mathfrak{s}_{i,t}^{V} = \left[V^{\min} - V_i^t\right]_+ + \left[V_i^t - V^{\max}\right]_+ \tag{40}$$

$$\mathfrak{s}_{ij,t}^{I} = \left[I_{ij}^t / I_{ij}^{lim} - 1\right]_+ \tag{41}$$

where $V_i^t$ denotes the voltage magnitude with limits $V^{\min}$ and $V^{\max}$, and the positive part function is $[x]_+ \triangleq \max(x, 0)$. Based on the violation severity, the hourly system-level voltage and current violation metrics are constructed. $S_t^{V,\text{tot}}$, $S_t^{V,\max}$ and $S_t^{V,\text{cnt}}$ represent the total severity, maximum severity and total count of nodal voltage violations at time $t$, respectively. The terms $S_t^{I,\text{tot}}$, $S_t^{I,\max}$ and $S_t^{I,\text{cnt}}$ denote the corresponding metrics for branch current violations between node $i$ and node $j$. For $f \in \{I, V\}$, the $u$-th violation event $E_u^f$ is defined as a maximal contiguous interval satisfying $S_t^{f,\text{cnt}} > 0$ bounded by start and end indices $\tau_{u,st}^f$ and $\tau_{u,e}^f$, respectively. Its corresponding duration is given by $d_u^f = \tau_{u,e}^f - \tau_{u,st}^f + 1$.

$$E_u^f = \left\{\forall t \in \mathcal{T}: \tau_{u,st}^f \le t \le \tau_{u,e}^f,\ S_t^{f,\text{cnt}} > 0\right\}, f \in \{I, V\} \tag{42}$$

The violation event feature vector is then defined as (43) including the duration, total severity, total count, maximum count and maximum severity within the violation event.

$$\boldsymbol{z}_u^f = \left[d_u^f, \sum\nolimits_{t \in E_u^f} S_t^{f,\text{tot}}, \sum\nolimits_{t \in E_u^f} S_t^{f,\text{cnt}}, \max_{t \in E_u^f} S_t^{f,\text{cnt}}, \max_{t \in E_u^f} S_t^{f,\max}\right]^T \tag{43}$$

For the high load-related event set, the $u$-th event is defined as a load peak-basin interval extracted from the annual load trajectory, with start, peak, and end indices denoted by $\tau_{u,st}^L$, $t_u^p$, and $\tau_{u,e}^L$, respectively. Its corresponding duration is given by $d_u^L = \tau_{u,e}^L - \tau_{u,st}^L + 1$. Each load event is characterized by its duration, peak load, peak prominence, accumulated basin energy, and maximum ramp within the interval. The corresponding load-event feature vector is defined as:

$$\boldsymbol{z}_u^L = \left[d_u^L, L_{t_u^p}, P_u^L, E_u^{L,\text{eng}}, R_u^{L,\max}\right]^T \in \mathbb{R}^5 \tag{44}$$

where $L_{t_u^p}$ is the peak load, $P_u^L$ denotes the peak prominence relative to the basin boundary, $E_u^{L,\text{eng}}$ denotes the accumulated load stress within the basin, and $R_u^{L,\max}$ denotes the maximum absolute load ramp during the event.

Let $\mathcal{K}^{\mathcal{F}}$ denote the set of all elementary events belonging to the set $\mathcal{F} \in \mathcal{M} = \{I, V, L\}$. To identify the most severe stress patterns, we evaluate events based on the feature vectors. For any two events $u, v \in \mathcal{K}^{\mathcal{F}}$, the event $v$ is determined to weakly dominate event $u$ denoted as $\boldsymbol{z}_v^{\mathcal{F}} \succcurlyeq \boldsymbol{z}_u^{\mathcal{F}}$ if it is no less severe in every feature dimension.

$$\boldsymbol{z}_v^{\mathcal{F}} \succcurlyeq \boldsymbol{z}_u^{\mathcal{F}} \Leftrightarrow z_{v,m}^{\mathcal{F}} \ge z_{u,m}^{\mathcal{F}}, \quad \forall m = 1, \dots, M_{\mathcal{F}} \tag{45}$$

where $M_{\mathcal{F}}$ is the dimensionality of the feature vector for set $\mathcal{F}$. Furthermore, the strict domination is defined as:

$$\boldsymbol{z}_v^{\mathcal{F}} \succ \boldsymbol{z}_u^{\mathcal{F}} \Leftrightarrow (\boldsymbol{z}_v^{\mathcal{F}} \succcurlyeq \boldsymbol{z}_u^{\mathcal{F}}, \forall m) \land (\exists m: \boldsymbol{z}_v^{\mathcal{F}} \succ \boldsymbol{z}_u^{\mathcal{F}}) \tag{46}$$

Governed by the dominance relations in (45)-(46), Pareto screening is applied within each event set to identify critical events, removing dominated events as shown below:

$$\widehat{\mathcal{K}}^{\mathcal{F}} = \{u \in \mathcal{K}^{\mathcal{F}} \mid \nexists v \in \mathcal{K}^{\mathcal{F}}: \boldsymbol{z}_v^{\mathcal{F}} \succ \boldsymbol{z}_u^{\mathcal{F}}\}, \mathcal{F} \in \mathcal{M} \tag{47}$$

Following the Pareto screening results, a set of retained events $\widehat{\mathcal{K}}^{\mathcal{F}}$ is obtained for each category $\mathcal{F} \in \mathcal{M}$. These retained events are then remapped onto the full annual timeline, and an hourly activation indicator is defined as:

$$a_t^{\mathcal{F}} = \begin{cases} 1, & \exists u \in \widehat{\mathcal{K}}^{\mathcal{F}} \text{ such that } t \in E_u^{\mathcal{F}}, \\ 0, & \text{otherwise,} \end{cases} \quad t \in \mathcal{T}, \mathcal{F} \in \mathcal{M} \tag{48}$$

### *A.2 Signature-Consistent Segment Construction*

For each time $t$, the corresponding active-type set $\Gamma_t$ and signature space $\mathcal{A}$ can be defined as:

$$\Gamma_t = \left\{\mathcal{F} \in \mathcal{M} \mid a_t^{\mathcal{F}} = 1\right\}, t \in \mathcal{T} \tag{49}$$

$$\mathcal{A} = 2^{\mathcal{M}} \backslash \{\emptyset\} = \{I,\ V,\ L,\ I \cap V,\ I \cap L,\ V \cap L,\ I \cap V \cap L\} \tag{50}$$

where $2^{\mathcal{M}}$ denotes the power set of $\mathcal{M}$ and for three categories of event types $\mathcal{M} = \{I, V, L\}$, $\mathcal{A}$ comprises a total of seven nonempty combinations.

Consecutive hours with the same signature label are merged into a continuous segment. Specifically, the $r$-th segment is defined as:

$$R_r = \left\{t \in \mathcal{T} \mid \tau_{r,st} \le t \le \tau_{r,e},\ \Gamma_t = \Gamma_r\right\} \tag{51}$$

where $\Gamma_r \in \mathcal{A}$ denotes the signature label of segment $R_r$, and the interval $[\tau_{r,st}, \tau_{r,e}]$ is maximal under hourly continuity. The segment duration is given by $d_r = \tau_{r,e} - \tau_{r,st} + 1$. The segment-level descriptive metrics are defined as follows:

$$\begin{cases} \bar{f}_r^{\text{tot}} = \dfrac{1}{d_r} \sum_{t \in R_r} S_t^{f,\text{tot}} & L_r^{\max} = \max\limits_{t \in R_r} L_t \\ \bar{f}_r^{\text{cnt}} = \dfrac{1}{d_r} \sum_{t \in R_r} S_t^{f,\text{cnt}}, & \bar{L}_r = \dfrac{1}{d_r} \sum_{t \in R_r} L_t \\ f_r^{\max} = \max\limits_{t \in R_r} S_t^{f,\max} & R_r^{L,\max} = \max\limits_{\tau_{r,s}+1 \le t \le \tau_{r,e}} |L_t - L_{t-1}| \end{cases} \tag{52}$$

For each retained segment $R_r$, a feature vector $\boldsymbol{x}_r$ dependent on the feature signature is constructed using the descriptors defined in (53). This mapping relationship can be concisely expressed as:

$$\boldsymbol{x}_r = \Phi(\Gamma_r) \tag{53}$$

The specific feature components of all seven nonempty signatures are summarized in Table 2. Upon the completion of segment construction, the Pareto screening process is invoked, executing independently within each feature category. To ensure fairness in evaluation, each segment is compared solely against peer segments sharing identical driving features, with the comparison based on their respective specific feature vectors. By filtering out dominated instances, this process yields a streamlined, non-dominated subset comprising representative high-stress periods.

Table 2. Combinations of seven nonempty signatures and feature vectors.

| Signature $\Gamma_r$ | Segment feature vector $x_r$ |
|---|---|
| $\{L\}$ | $\left[d_r, L_r^{\max}, \bar{L}_r, R_r^{L,\max}\right]^{\mathrm{T}}$ |
| $\{I\}$ | $\left[d_r, I_r^{\max}, \bar{I}_r^{\text{tot}}, \bar{I}_r^{\text{cnt}}\right]^{\mathrm{T}}$ |
| $\{V\}$ | $\left[d_r, V_r^{\max}, \bar{V}_r^{\text{tot}}, \bar{V}_r^{\text{cnt}}\right]^{\mathrm{T}}$ |
| $\{V \cap L\}$ | $\left[d_r, V_r^{\max}, \bar{V}_r^{\text{tot}}, L_r^{\max}, R_r^{L,\max}\right]^{\mathrm{T}}$ |
| $\{I \cap L\}$ | $\left[d_r, I_r^{\max}, \bar{I}_r^{\text{tot}}, L_r^{\max}, R_r^{L,\max}\right]^{\mathrm{T}}$ |
| $\{I \cap V\}$ | $\left[d_r, I_r^{\max}, \bar{I}_r^{\text{tot}}, V_r^{\max}, \bar{V}_r^{\text{tot}}\right]^{\mathrm{T}}$ |
| $\{I \cap V \cap L\}$ | $\left[d_r, I_r^{\max}, \bar{I}_r^{\text{tot}}, V_r^{\max}, \bar{V}_r^{\text{tot}}, L_r^{\max}, R_r^{L,\max}\right]^{\mathrm{T}}$ |

Although Pareto screening removes dominated stress segments, the retained set may still include highly similar candidates. Therefore, a K-medoids compression step is applied within each surviving signature class to retain a compact subset of representative historical segments while preserving stress-pattern diversity. Each retained segment is mapped back to the retained primitive events that overlap it. For a retained segment $R_r$, define its original-event set as:

$$\mathcal{P}_r = \left\{ E_u^{\mathcal{F}} \in \widehat{\mathcal{K}}^{\mathcal{F}} \mid E_u^{\mathcal{F}} \cap R_r \neq \emptyset,\ \mathcal{F} \in \Gamma_r \right\}, R_r \in R^{\text{rep}} \tag{54}$$

where $\Gamma_r \subseteq \{I, V, L\}$ denotes the active-type set associated with segment $R_r$.

The final planning horizon associated with $R_r$ is recovered as the continuous interval covering all original events:

$$\tau_{r,st}^{W} = \min_{E_u^{\mathcal{F}} \in \mathcal{P}_r} \tau_{k,st}^{\mathcal{F}}, \tau_{r,e}^{W} = \max_{E_u^{\mathcal{F}} \in \mathcal{P}_r} \tau_{k,e}^{\mathcal{F}} \tag{55}$$

$$W_r = \{t \in \mathcal{T}: \tau_{r,st}^{W} \leq t \leq \tau_{r,e}^{W}\} \tag{56}$$

where if multiple retained segments correspond to the same recovery interval, the identical horizons are merged. The resulting set $W_r$ constitutes the final set of planning horizons.

*A.3 Cross-Horizon Validation*

The finalized set of planning horizons is subsequently evaluated within the T-OEP model to generate distinct candidate installation schemes. These candidate schemes are cross-validated against all extracted horizons. The feasibility indicator is then defined as:

$$\delta_{m,n} = \begin{cases} 1, \text{if solution of } m \text{ is feasible on } n \\ 0, \text{otherwise} \end{cases} \quad m, n = 1,2, \dots, w \tag{57}$$

where $w$ represents the total number of final planning horizons, and the element $\delta_{m,n}$ represents the feasibility status when the optimized investment scheme obtained from horizon $m$ is validated against the stress conditions of horizon $n$.

Based on the feasibility indicator, define the set of fully feasible source horizons as:

$$\mathcal{H}^{\text{all}} = \{m \in \{1,2, \dots, w\} \mid \delta_{m,n} = 1, \forall n = 1,2, \dots, w\} \tag{58}$$

The final planning period is selected as the source horizon, whose planning solution remains feasible across all target horizons and yields the minimum objective value for the source horizon.

$$W^{\star} = W_{m^{\star}}, m^{\star} \in \arg \min_{m \in \mathcal{H}^{\text{all}}} \mathcal{J}_m \tag{59}$$

where $\mathcal{J}_m$ represents the optimal objective value of objective (39) obtained from horizon $m$.

*B. Adaptive Spatial Targeting*

The second stage of the proposed STAT framework, adaptive spatial targeting, determines the reduced set of candidate installation nodes for BESS and SC based on the sensitivity metrics, load metrics and topological metrics. The STAT-AST method achieves dimensionality reduction of the solution space in planning tasks by adaptively constructing a pool of stress-relevant and electrically influential candidate buses through the screening of feature vectors.

*B.1 Device-Specific Feature Construction*

For each candidate node $n \in \mathcal{N}$, the cumulative voltage and current stress weights are aggregated over the operational horizon, as formulated in (60)-(61).

$$\omega_i^{V} = \sum_{t \in \mathcal{T}} \mathfrak{s}_{i,t}^{V}, \forall i \in \mathcal{N} \tag{60}$$

$$\omega_{ij}^{I} = \sum_{t \in \mathcal{T}} \mathfrak{s}_{ij,t}^{I}, \forall (i,j) \in \mathcal{L} \tag{61}$$

The voltage sensitivities are derived analytically from a LinDistFlow approximation using the shared electrical path between the observed and injection buses. The branch-current sensitivity is obtained from the local relationship among branch active power, reactive power, voltage, and current. Based on these sensitivities and the voltage and current violation weights, three stress-informed sensitivity metrics are constructed as:

$$\begin{cases} S_n^{VP} = \sum_{i \in \mathcal{N}} \omega_i^{V} |\partial V_i / \partial P_n| \\ S_n^{VQ} = \sum_{i \in \mathcal{N}} \omega_i^{V} |\partial V_i / \partial Q_n| \quad , \forall n \in \mathcal{N} \\ S_n^{IP} = \sum_{(i,j) \in \mathcal{L}} \omega_{ij}^{I} |\partial I_{ij} / \partial P_n| \end{cases} \tag{62}$$

where $S_n^{VP}$ and $S_n^{VQ}$ denote the voltage sensitivity indices associated with active and reactive power injections at bus $n$, and $S_n^{IP}$ represents the stress-weighted branch current sensitivity magnitude associated with active power injections at bus $n$. $V_i$ is the voltage magnitude at bus $i$, $I_{ij}$ is the current magnitude between node $i$ and $j$, and $P_n$ and $Q_n$ denote the active and reactive power injections at candidate bus $n$.

For BESS, the candidate node screening is oriented toward active-power support and upstream-current relief. In addition to the $S_n^{VP}$ and $S_n^{IP}$, the local peak demand $P_n^{\max}$ and local maximum ramp $r_n^{P}$ are incorporated as:

$$P_n^{\max} = \max_{t \in \mathcal{T}} P_{\text{load},n}^{t} \tag{63}$$

$$r_n^{P} = \max_{t \in \mathcal{T} \setminus \{t_0\}} \left| P_{\text{load},n}^{t} - P_{\text{load},n}^{t-1} \right| \tag{64}$$

The feature vector of candidate nodes for BESS installation is defined in (65). This vector prioritizes the selection of nodes that are capable of effectively mitigating terminal voltage dips while simultaneously facing significant local peak load or load ramp rate pressures.

$$\boldsymbol{z}_n^{\text{BESS}} = [S_n^{VP}, S_n^{IP}, P_n^{\max}, r_n^{P}]^{T} \in \mathbb{R}^4, \forall n \in \mathcal{N} \tag{65}$$

The screening of candidate nodes for SC installation primarily focuses on providing local voltage support and reducing upstream reactive-power flow. Alongside the metric $S_n^{VQ}$, the evaluation incorporates spatial, topological, and reactive-load metrics: the support focus index $\phi_n$, which quantifies the degree of electrical coupling between candidate node $n$ and the voltage-constrained regions; and the peak downstream reactive power demand $q_n^{\text{sub,max}}$, which characterizes the aggregate reactive power support required by the topological subtree rooted at node $n$; and the local reactive peak load $q_n^{\text{loc,max}}$ and local reactive concentration $\rho_n^{Q}$.

$$\phi_n = \sum_{i\in\mathcal{N}} [\omega_i^V/(1 + d^{\mathrm{el}}(i,n))] \quad (66)$$

$$q_n^{\mathrm{sub,max}} = \max_{t\in\mathcal{T}} \sum_{i\in\mathcal{D}(n)} Q_{\mathrm{load},i}^t \quad (67)$$

$$q_n^{\mathrm{loc,max}} = \max_{t\in\mathcal{T}} Q_{\mathrm{load},n}^t \quad (68)$$

$$\rho_n^Q = q_n^{\mathrm{loc,max}}/(q_n^{\mathrm{sub,max}} + \epsilon) \quad (69)$$

where $d^{\mathrm{el}}(i,n)$ represents the electrical distance between node $i$ and $n$, and $\epsilon$ is a small positive constant to avoid division by zero. Therefore, the feature vector of candidate nodes for SC installation is defined as $\boldsymbol{z}_n^{\mathrm{SC}}$ in (70) which is primarily used to identify nodes that possess robust reactive voltage support capabilities, are electrically proximate to voltage-stressed regions, and are capable of meeting the substantial reactive power demands of downstream areas.

$$\boldsymbol{z}_n^{\mathrm{SC}} = [S_n^{VQ}, \phi_n, q_n^{\mathrm{sub,max}}, q_n^{\mathrm{loc,max}}, \rho_n^Q]^T \in \mathbb{R}^5, \forall n \in \mathcal{N} \quad (70)$$

*B.2 Candidate Bus Screening*

The Pareto screening method defined above is applied separately to the feature set reduction of BESS and SC nodes. The non-dominated nodes retained after screening constitute the reduced sets of candidate nodes for these two device types, which will be used in the subsequent planning model.

*C. STAT-Enabled T-OEP Implementation and Validation*

The STAT-enabled T-OEP is implemented through three stages: critical-horizon planning, cross-horizon validation, and full-year operational validation.

First, each critical horizon extracted by the STAT-TCA method is treated as a source horizon to generate a candidate investment scheme. The T-OEP model jointly optimizes line upgrades, BESS locations and capacities, inverter ratings, SC installations, and operational variables over the truncated horizon. Because the horizon boundaries are determined by temporal screening rather than by a complete operating cycle, the cyclic SoC constraint in (33) is relaxed only at this planning stage. The initial BESS energy is optimized within the admissible range in (31), while the sequential energy dynamics in (32) remain enforced.

Second, the investment decisions obtained from each source horizon are fixed and validated on all retained target horizons. Only operational variables, including power flows, BESS dispatch, SC output, and substation injections, are re-optimized. During cross-horizon validation, the cyclic SoC constraint in (33) is restored and enforced together with the BESS energy, voltage, and branch-current limits. A source-horizon scheme is considered transferable only if it remains feasible for all target horizons. If multiple transferable schemes exist, the one with the minimum investment cost is selected. If no individual source horizon remains feasible across all target horizons, the cross-horizon validation results are used to identify horizons representing distinct feeder-level stress conditions. Redundant horizons are removed, and the smallest complementary set covering the observed validation failures is retained for the final composite-horizon T-OEP.

Finally, the selected investment scheme is fixed and evaluated over the full chronological annual trajectory. The BESS energy dynamics and cyclic SoC constraint are enforced, and all hourly voltage, current, BESS, SC, and inverter limits are checked. Therefore, any under-sized investment caused by the relaxed SoC boundary in the planning stage is rejected during cross-horizon or full-year validation, and branch-loss minimization is used during validation to promote SOCP tightness, which is verified through residual checks.

## IV. Case Studies

*A. Test Systems and Implementation Settings*

The test systems include standard IEEE 33-bus system and 240-bus test system [29]. The base apparent power and base voltage of test systems are set to 10 MVA and 13.8 kV. Based on the empirical dataset, we simulate the effects of demand growth combined with increased EV penetration. The EV charging profiles used in the case studies are generated based on the annual charging-scenario modeling framework [2]. In that study, historical EV-user charging behavior was analyzed to construct stochastic annual charging scenarios, and a Monte Carlo simulation was used to generate yearly charging profiles for four Level-2 charger categories. The four charger classes are characterized by charging-rate ranges of 4-7 kW, 7-11.2 kW, 11.2-15 kW, and above 15 kW, respectively. Following the observed charger-class distribution, the EV charging loads in this study are randomly assigned to all users at each node according to the proportions of 15%, 55%, 20%, and 10% for the four charger categories. The numbers of EV users in the 33-bus and 240-bus system are 519 and 1,120, respectively. These sampled charging profiles are then superimposed on the baseline nodal load to represent spatially distributed EV charging demand across the feeder.

The nodal installation capacity of BESS is limited to 2 MWh with a 10-year lifespan. The unit costs are set to \$241/kWh for the energy-related capacity cost $C_{\mathrm{BESS}}^{\mathrm{eng}}$ and \$310/kW for power-related rating cost $C_{\mathrm{BESS}}^{\mathrm{pow}}$ [30]. Each SC bank has a discrete capacity of 150 KVAr and a 20-year lifespan. Installations are capped at 8 banks per node. The cost includes a fixed installation fee $C_{\mathrm{SC}}^{\mathrm{fix}}$ of \$3,000 per site and a step cost $C_{\mathrm{SC}}^{\mathrm{step}}$ of \$4,250 per bank. The capital expenditure for LR varies between \$150,000/km and \$470,000/km depending on the specific cable type, assuming a 30-year service life. The annual discount rate is assumed to be 5% for all annualized cost calculations. The MIP optimality gap is set to 0.5% of the planning model. The optimization calculation was conducted using Gurobi Optimizer v12.0.3 and the computational environment utilized a 12th Gen Intel Core™ i7-12700 CPU. The model was implemented in Python 3.9.

*B. IEEE 33-Bus System: Temporal and Spatial Benchmarking*

The proposed STAT-TCA method operates at an hourly resolution for event segmentation and segment processing. As a result, the extracted planning horizons are continuous stress episodes rather than fixed 24-hour blocks. This data-driven selection of contiguous time horizons avoids artificially interrupting critical operating trajectories and prolonged stress periods, which is a common limitation of conventional day-based partitioning. The candidate critical planning horizons and associated signatures are presented in Table 3. The final critical planning horizon selected by the STAT framework is from 07-06 06:00 to 07-07 05:00.

Table 3 summarizes the candidate horizons retained by STAT-TCA method, the corresponding planning results, and the cross-horizon validation outcomes in the 33-bus system. The retained horizons cover different combinations of current-overload, voltage-violation, and load-stress signatures,

Table 3. Candidate Horizons, Planning Results, and Cross-Horizon Validation in the 33-Bus System.

| Horizon | Start time | End time | Signature | LR (A) | BESS (kWh) | SC (kVAr) | Annualized Cost | Runtime | Failed Targets |
|---|---|---|---|---|---|---|---|---|---|
| $W_1$ | 06-13 18:00 | 06-14 05:00 | $\{I \cap L\}$ | / | 9/8,101 | 5/3,900 | $458,026 | 45 s | $W_2$, $W_3$, $W_4$ |
| $W_2$ | 07-06 06:00 | 07-07 05:00 | $\{I \cap V \cap L\}$ | 1-2/650 | 9/11,570 | 4/3,750 | $680,614 | 222 s | / |
| $W_3$ | 07-09 06:00 | 07-10 05:00 | $\{I \cap V \cap L\}$ | / | 9/8,140 | 6/5,100 | $463,149 | 82 s | $W_2$, $W_4$ |
| $W_4$ | 07-17 06:00 | 07-19 04:00 | $\{V \cap L\}$ | 2-3/650 | 8/8,327 | 7/4,500 | $639,898 | 1187 s | $W_2$ |
| $W_5$ | 07-19 06:00 | 07-22 03:00 | $\{V \cap L\}$ | / | 9/8,041 | 6/4,350 | $455,987 | 819 s | $W_1$ ,$W_2$, $W_3$, $W_4$ |

Note: "Failed targets" indicate target horizons where voltage and/or current violations remain after fixing the source-horizon investment decisions.

indicating that the operational violations are driven by distinct but partially overlapping temporal stress patterns. The LR column reports the upgraded branch and its selected ampacity, while the BESS and SC columns report the number of installed devices and their total capacities. The annualized cost and runtime are also listed for each planning horizon.

After each source-horizon planning problem is solved, its investment decisions are fixed and only the operational variables are re-optimized on the other retained target horizons. A source-horizon scheme is regarded as transferable only if it eliminates both voltage and current violations across all target horizons. As shown in Table 3, $W_1$, $W_3$, $W_4$ and $W_5$ all fail at least one target horizon, even though several of them yield lower source-horizon investment costs. In contrast, the $W_2$ derived scheme has no failed target horizon and is the only candidate that passes the cross-horizon validation. Accordingly, $W_2$is selected as the final STAT-TCA planning horizon. Its annualized cost of $680,614 is higher than those of some rejected candidates, but this higher cost reflects the investment required to maintain transferable feasibility across all retained critical horizons. In this case, directly selecting the minimum-cost horizon without cross-horizon validation would produce an investment scheme that fails under other retained stress conditions.

For comparison, several benchmark horizon-selection methods are considered. The peak-violation day and week are identified by ranking candidate windows according to a weighted combination of violation count, severity, and duration. The peak-load day and week contain the annual maximum system demand. The CDF-based baselines select the day or week whose load distribution most closely matches the annual empirical distribution. The K-medoids baseline clusters standardized window features describing voltage violations, current overloads, load level, and load ramp, and selects the highest-stress medoid. The STAT-TCA method without the load-stress event family, denoted as STAT-TCA w/o L, is also included to assess the contribution of load information. All methods use the same T-OEP formulation, candidate sets, investment parameters, and solver settings.

Table 4 compares the resulting planning schemes and their validation performance. The investment decisions obtained from each selected horizon are fixed and evaluated under the prescribed validation procedure. STAT-TCA produces a feasible scheme with an annualized cost of $680,614 and a runtime of 222.07 s. The peak-violation week also passes validation and yields a comparable investment plan, but the runtime increases to 26,519.75 s. Removing the load-stress family expands the recovered STAT-TCA horizon from 23 h to 92 h and increases the runtime to 13,755.91 s, indicating that load information helps refine the temporal boundaries of the critical stress episode. In contrast, the peak-violation day, peak-load, CDF, and K-medoids baselines do not pass validation in this case, showing that severity-based, demand-based, distribution-based, or clustering-based selection alone does not necessarily yield transferable investment decisions. The full-year formulation was not solved within the prescribed time limit.

As shown in Fig. 2(a), both the Top-K baseline and the proposed STAT-AST method reduce the Full-node candidate pool to 17 BESS candidate buses and 22 SC candidate buses, corresponding to reductions of 47% and 31%, respectively. Under the same candidate-set size, STAT-AST achieves shorter solver runtime than the Top-K method. As illustrated in Fig. 2(b), the solver runtime is reduced from 222.07 s in the Full-node case to 88.62 s with STAT-AST, corresponding to a 2.51-fold speedup. Compared with the Top-K baseline, STAT-AST further reduces the runtime from 121.39 s to 88.62 s, yielding a 1.37-fold speedup. These results demonstrate that the proposed device-specific spatial targeting can effectively reduce the candidate search space while retaining a more computationally efficient candidate set for the optimization.

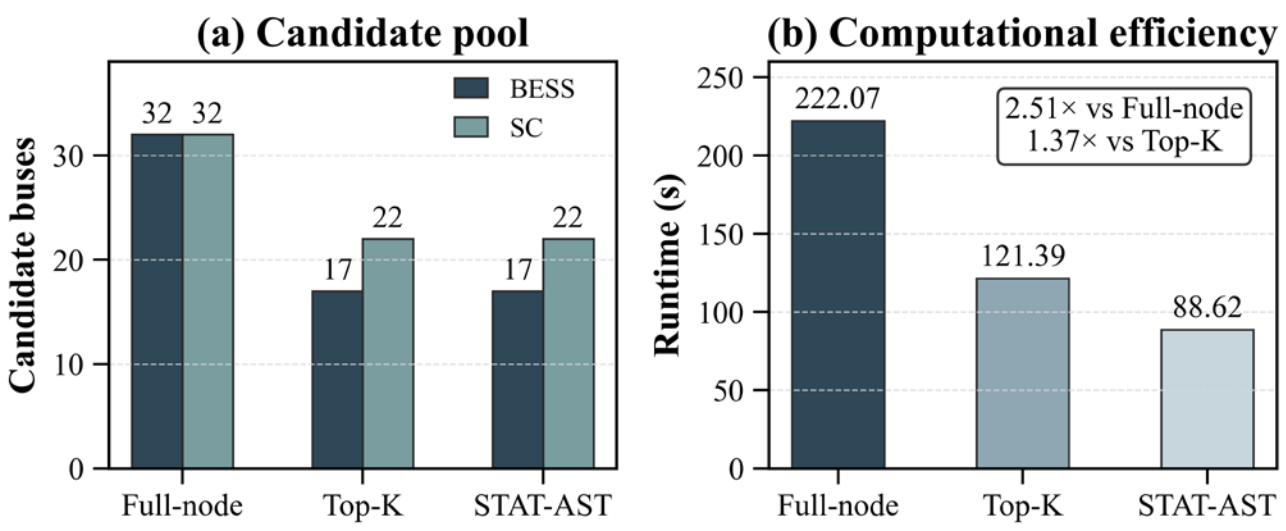

Fig. 2. Performance comparison among the Full-node, Top-K, and STAT-AST candidate-selection methods: (a) candidate pool reduction, and (b) computational efficiency.

Despite the reduced candidate space, Table 5 shows that STAT-AST preserves planning performance comparable to the Full-node benchmark. The deviations in BESS capacity and annualized cost are only 0.01% and 0.009%, respectively, while the installed SC capacity remains identical. Although the reported STAT-AST incumbent has a slightly lower annualized cost than the Full-node incumbent, this difference is well within the 0.5% MIP optimality tolerance and therefore does not indicate that the reduced candidate set outperforms the Full-node formulation. Instead, the two solutions are regarded as practically equivalent within the reported solver tolerance. Under the same candidate-set size, STAT-AST also reproduces the Full-node installed SC capacity. These results indicate that the device-specific spatial targeting reduces the candidate space and solver runtime while matching the Full-node planning benchmark within the reported solver tolerance.

Table 5. Planning Performance Comparison under Different Spatial Candidate-Selection Strategies.

| Method | BESS (MWh) | BESS Dev. | SC (kVAr) | Annualized Cost | Cost Dev. |
|---|---|---|---|---|---|
| Full-node | 11.571 | - | 3,750 | $680,674 | - |
| Top-K | 11.579 | 0.07% | 3,600 | $680,751 | 0.011% |
| STAT-AST | 11.570 | 0.01% | 3,750 | $680,614 | 0.009% |

The detailed configurations of this optimized scheme are visually mapped onto the 33-bus system topology in Fig. 3 and subsequently evaluated through full-year operational validation.

Table 4. Performance comparison of different planning-horizon selection methods.

| Method | Start time | End time | LR (A) | BESS (kWh) | SC (kVAr) | Annualized Cost | Runtime | Validation |
|---|---|---|---|---|---|---|---|---|
| STAT-TCA | 07-06 06:00 | 07-07 05:00 | 1-2/650 | 9/11,570 | 4/3,750 | $680,614 | 222 s | Pass |
| STAT-TCA w/o *L* | 07-04 08:00 | 07-08 04:00 | 1-2/650 | 9/11,571 | 4/3,750 | $680,634 | 13,756 s | Pass |
| Peak-violation week | 07-06 00:00 | 07-13 00:00 | 1-2/650 | 9/11,571 | 4/3,750 | $680,630 | 26,520 s | Pass |
| Peak-violation day | 07-06 00:00 | 07-07 00:00 | 1-2/650 | 9/7,924 | 6/4,200 | $480,489 | 135 s | Fail |
| Peak-load week | 07-15 00:00 | 07-22 00:00 | 2-3/650 | 9/8,068 | 7/4,500 | $639,730 | 39,112 s | Fail |
| Peak-load day | 07-18 00:00 | 07-19 00:00 | 1-2/650 | 9/5,014 | 7/5,850 | $323,579 | 116 s | Fail |
| CDF week | 06-13 00:00 | 06-20 00:00 | / | 8/8,324 | 6/4,500 | $457,810 | 21,927 s | Fail |
| CDF day | 07-08 00:00 | 07-09 00:00 | / | 5/942 | 7/7,050 | $69,804 | 79 s | Fail |
| K-Medoids day | 07-19 01:00 | 07-20 01:00 | / | 9/6,365 | 6/4,800 | $364,351 | 96 s | Fail |
| Full-year planning | 01-01 00:00 | 12-31 23:00 | No incumbent solution found within 172,800 s | | | | | |

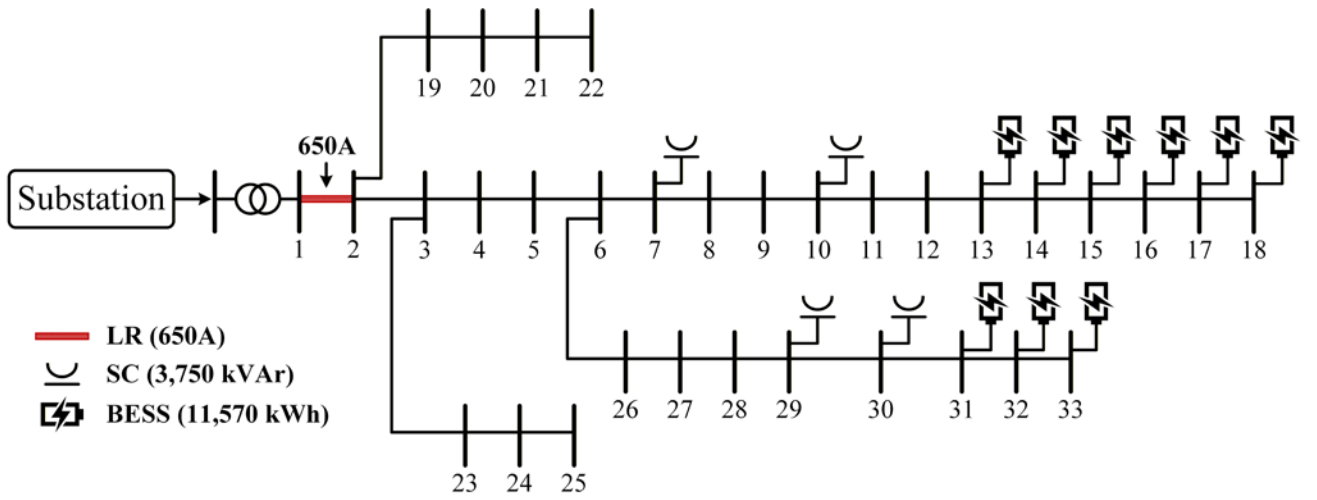

Fig. 3. Optimized deployment of the proposed planning solution in the IEEE 33-bus system.

As shown in the box plots in Fig. 4, the original EV-induced violation case resulted in multiple bus voltages falling below the lower limit of 0.95 p.u. After applying the STAT-based optimal planning solution, the voltage profiles were significantly improved and maintained within the acceptable operating range.

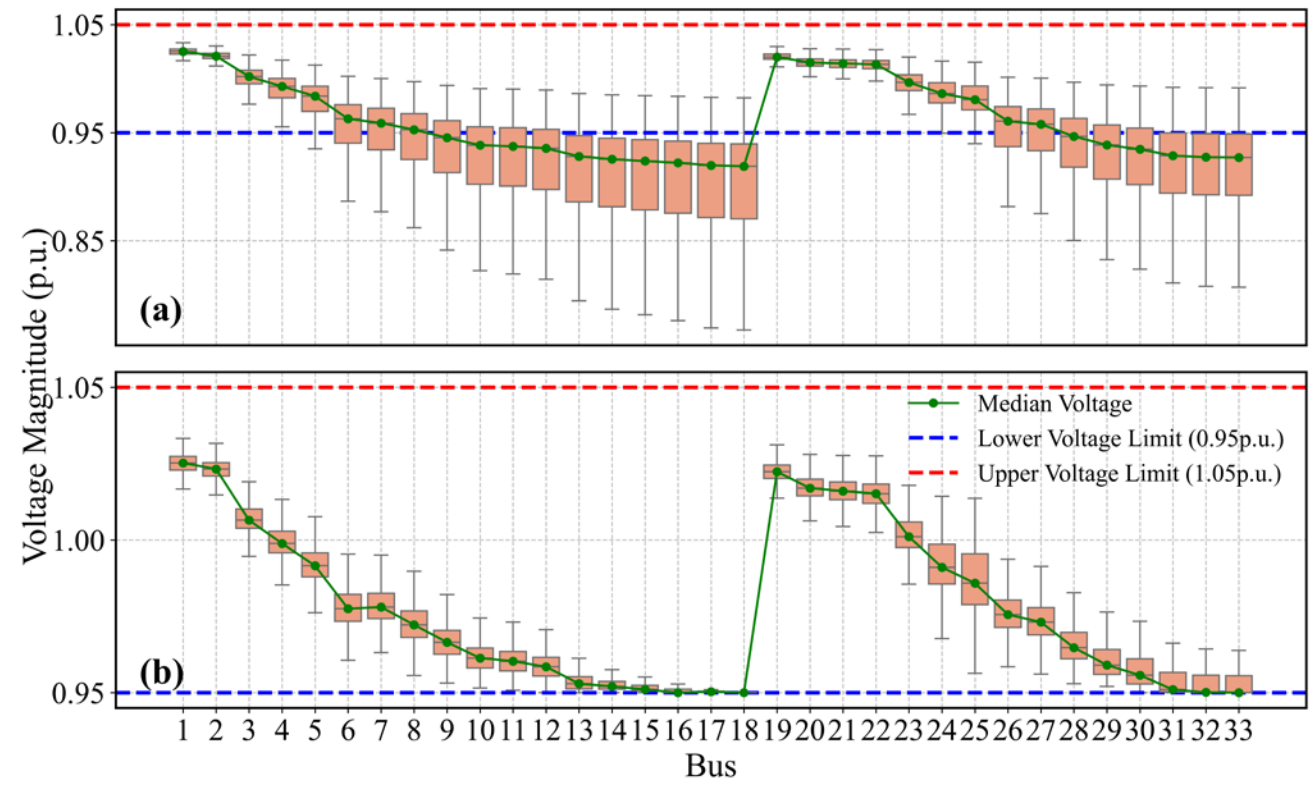

Fig. 4. Annual voltage distribution before and after implementing the STAT-based optimized planning solution in 33-bus system: (a) original EV-induced violation case and (b) validated STAT-based optimal solution.

### *C. The 240-Bus System: Larger-System Application*

Based on the VA model results, a total of 141 nodes and 24 branches exhibit voltage and current limit violations in the 240-bus system. This case is used to further evaluate the scalability of the proposed STAT-enabled planning framework on a larger multi-feeder distribution network. Under STAT-TCA, the retained critical horizons exhibit feeder-dependent stress patterns. The cross-horizon validation indicates that no single critical horizon fully represents all feeder-level stress conditions. But two critical horizons provide complementary coverage of the dominant feeder-level violations where Horizon I captures the dominant stress conditions in Feeder B, whereas Horizon II captures the complementary violations in Feeder C. Therefore, the final planning solution is derived from a composite set of Horizons I and II and then verified through full-year operational validation.

As summarized in Table 6, Horizon I spans from 07-08 00:00 to 07-14 00:00, while Horizon II spans from 07-16 00:00 to 07-19 04:00. STAT-AST reduces the spatial candidate set to 36 BESS candidate buses and 63 SC candidate buses out of 239 buses, corresponding to 85% and 74% reductions, respectively. Based on the composite critical-horizon set, the final validated solution installs 33 BESS units with a total energy capacity of 20,577 kWh, one SC installation with a total capacity of 1,200 kVAr, and 19 LR upgrades along the main stressed corridors. The final annualized cost is $1,225,390, and the critical-horizon planning runtime is 201,602.9 s. The resulting deployment passes the full-year operational validation, confirming the applicability of the proposed framework to the tested larger multi-feeder system while maintaining year-round feasibility.

Table 6. Planning and Validation Summary for the 240-Bus System.

| Item | Result |
|---|---|
| **Feeder-dependent critical horizons** | |
| Critical Horizon I-Feeder B | 07-08 00:00~07-14 00:00 |
| Horizon I Annualized Cost | $1,033,586 |
| Horizon I Planning Runtime | 28,801 s |
| Critical Horizon II-Feeder C | 07-16 00:00~07-19 04:00 |
| Horizon II Annualized Cost | $1,222,654 |
| Horizon II Planning Runtime | 172,802 s |
| **Final validated planning solution** | |
| Final critical-horizon source | Composite set of Horizons I, II |
| Installed BESS sites /capacities | 33 / 20,577 kWh |
| Installed SC sites /capacities | 1 / 1,200 kVAr |
| LR upgrades / main corridors | 19 / Branches: 1-2002, 1-3082 |
| Main upgraded breakers / ampacity | Breaker / 630, 1250 A |
| Main upgraded cables / ampacity | Cable / 443-650 A |
| Final annualized cost | $1,225,390 |
| Critical-horizon planning runtime | 201,603 s |
| Full-year operational validation | Pass |
| **Candidate reduction by STAT-AST** | |
| BESS candidate set | 36 / 239 (85% reduction) |
| SC candidate set | 63 / 239 (74% reduction) |

Table 7 summarizes the year-round voltage and current violation mitigation performance in both test systems. The arrows indicate the change from the original EV-induced violation case to the validated planning solution. In the 33-bus system, the minimum voltage is improved from 0.72 p.u. to 0.95 p.u., while the annual voltage-violation bus-hours and current-violation branch-hours are reduced from 120,614 to 0 and from 934 to 0, respectively. In the 240-bus system, the proposed solution eliminates violations across 141 initially violated buses and 24 overloaded branches, reducing the annual voltage-violation bus-hours from 102,163 to 0 and the current-violation branch-hours from 45,331 to 0. The maximum line loading is reduced from 137.9% to 81.87% in the 33-bus system and from 269.4% to 98.51% in the 240-bus system, confirming that the final validated deployments satisfy both voltage and thermal constraints over the full annual horizon. Here, V-violation bus-hours and I-violation branch-hours denote the cumulative numbers of bus–hour and branch–hour pairs exceeding the corresponding operating limits over the annual horizon. The BESS SoC trajectories are also checked during the full-year operational validation, and all installed BESS units remain within their admissible SoC limits while satisfying the cyclic end-condition.

Table 7. Full-Year Operational Violation Summary in the 33-Bus and 240-Bus Systems.

| System | Min. V (p.u.) | V-violation buses | V-violation bus-hours | Max line loading | I-violation branches | I-violation branch-hours |
|---|---|---|---|---|---|---|
| 33-bus | 0.72→0.95 | 25→0 | 120,614→0 | 137.9%→81.87% | 2→0 | 934→0 |
| 240-bus | 0.91→0.99 | 141→0 | 102,163→0 | 269.4%→98.51% | 24→0 | 45,331→0 |

## V. CONCLUSIONS

This paper proposed an OEP model to mitigate EV-induced voltage and thermal violations through coordinated LR, BESS, and SC investments. The proposed STAT framework reduces the temporal and spatial search spaces by identifying transferable critical planning horizons and device-specific candidate buses. In the 33-bus system, the STAT-TCA provided a compact horizon that outperformed the tested temporal baselines in computational efficiency, while STAT-AST matched the Full-node planning benchmark within the reported solver tolerance using reduced candidate sets. The 240-bus case demonstrated the applicability of the framework to a large-scale multi-feeder distribution system. Full-year operational validation confirmed voltage, current, and BESS operational feasibility for the tested EV scenarios. The results indicate that STAT framework improves the tractability of joint distribution expansion planning while preserving fidelity to the solvable benchmarks and annual feasibility